\documentclass[conference]{IEEEtran}
\IEEEoverridecommandlockouts
\usepackage{epsfig}
\usepackage{amsmath}
\usepackage{theorem}
\usepackage{amsmath}
\usepackage{mathrsfs}
\usepackage{amsfonts}
\usepackage{amssymb}
\usepackage{mathrsfs}
\usepackage{euscript}
\usepackage{graphicx}
\usepackage{multirow}
\usepackage{cite}
\usepackage{url}
\usepackage{color}

\newtheorem{proposition}{Proposition}

\newtheorem{remark}{Remark}

\def\BibTeX{{\rm B\kern-.05em{\sc i\kern-.025em b}\kern-.08em
    T\kern-.1667em\lower.7ex\hbox{E}\kern-.125emX}}
\begin{document}

\title{Asymptotic Close To Optimal Joint Resource Allocation and Power Control in the Uplink of Two-cell Networks
}
\author{
\IEEEauthorblockN{Ata Khalili, Soroush Akhlaghi and Meysam Mirzaee\\}
\IEEEauthorblockA{Electrical Engineering Department, Shahed University, Tehran, Iran\\
Emails: \{a.khalili,akhlaghi,me.mirzaee\}@shahed.ac.ir
}
}

\maketitle

\begin{abstract}
In this paper, we investigate joint resource allocation and power control mechanisms for two-cell networks, where each cell has some sub-channels which should be allocated to some users. The main goal persuaded in the current work is finding the best power and sub-channel assignment strategies so that the associated sum-rate of network is maximized, while a minimum rate constraint is maintained by each user. The underlying optimization problem is a highly non-convex mixed integer and non-linear problem which does not yield a trivial solution. In this regard, to tackle the problem, using an approximate function which is quite tight at moderate to high signal to interference plus noise ratio (SINR) region, the problem is divided into two disjoint sub-channel assignment and power allocation problems. It is shown that having fixed the allocated power of each user, the sub-channel assignment can be thought as a well-known assignment problem which can be effectively solved using the so-called Hungarian method. Then, the power allocation is analytically derived. Furthermore, it is shown that the power can be chosen from two extremal points of the maximum available power or the minimum power satisfying the rate constraint. Numerical results demonstrate the superiority of the proposed approach over the random selection strategy as well as the method proposed in~\cite{b3} which is regarded as the best known method addressed in the literature.
\end{abstract}


\section{Introduction}
Channel assignment and power allocation are an integral part of resource allocation in wireless networks, where the objective is to properly allocate the available bandwidth as well as the power of each user such that a performance criterion is maintained.

Recently, orthogonal frequency division multiple access (OFDMA) has received considerable attention due to its ability to divide the available bandwidth into some orthogonal sub-channels which can be simultaneously assigned to some users throughout the same symbol time period. OFDMA technique has been widely adopted in broadband wireless communications over the last decade, due to its flexibility in resource allocation.
Moreover, since the available bandwidth is divided into a number of orthogonal sub-channels, this can effectively mitigate the impact of frequency selective fading~\cite{b1,b2,b3,b4,b5,b6,b7}. It is worth mentioning that as users inside a cell make use of orthogonal sub-channels, the intra-cell interference is simply avoided in such networks . However, inter-cell interference has yet to be addressed~\cite{b2,b3,b4,b5,b6,b7,b8,b9,b10,b12}. It should be noted that due to the stringent power constraint associated with mobile equipment and the interference pattern, the impact of uplink scheduling and power control is more significant than that of the Downlink.

In this regard, a plethora of works are devoted to explore effective ways of assigning sub-channels as well as allocating power so that a performance function is optimized ~\cite{b3,b4,b5,b6}. For instance,~\cite{b3} investigates joint sub-channel assignment and power control in terms of maximizing the throughput in an uplink OFDMA network, where by adding a penalty term to the original problem, it is shown that the underlying mixed integer non-linear problem can be converted into a Difference of two Convex (DC) problem, where an approximate solution is devised.

The authors in~\cite{b4} maximized a weight sum rate in the uplink and formulates the joint scheduling and power control problem as a non-convex mixed integer nonlinear problem. The main contribution of ~\cite{b4} is a novel problem reformulation based on sum-of-ratios programming and a distributed iterative algorithm based on a subsequent quadratic transformation.In \cite{b7}the joint sub-channel assignment and power control problems in a cellular network with the objective of enhancing the quality of-service is studied. Accordingly, this problem is tackled in two steps. First, it attempts to assign the sub-channels assuming all users make use of an equal power. Then, the power of each user is optimized for the assigned channels. This problem has a poor performance as compared to\cite{b3}.
%~\cite{b5} proposed the series of problem reformulations based on the fractional programming and recursively solved the resulting convex approximation to find a local optimum of the proportionally fair resource allocation problem.

In this paper, we consider joint power and sub-channel allocation in the uplink direction of OFDMA network. The optimization problem is a highly non-convex mixed integer non-linear problem. The suboptimal power and resource allocation policy can be used by solving the considered problem via an optimization approach based on the sub-gradient method as in~\cite{b7} . But, we propose a low-complexity suboptimal algorithm based on the Hungarian method and it is shown that its results is close to optimal in high SNR regions. By other words, the main problem is divided into two step optimization problem, such as~\cite{b4,b5,b6,b7}. At first, we certain the sub-channel assignment for fix power assumption based on the Hungarian method. Then the optimal transmit power of users are obtained for each sub-channel in closed form solution. It is shown that the computational complexity of this method is much less than previous works and also its given throughput is close to optimal in high SNR region. 
\section{SYSTEM MODEL AND PROBLEM FORMULATION}
%\subsection{}
We consider the uplink channel of a two-cell OFDMA network with N available sub-channels and M users in each cell. The uplink channel strength value $h_{mj}^{nk}$ from the $m^{th}$ user in the $j^{th}$ cell to the BS of the $k^{th}$ cell in the $n^{th}$ sub-channel is assumed to be exponentially distributed (the channel gain is Rayleigh flat fading). Let $p_{mj}^{n}$ and $x_{mj}^{n}$ denote, respectively, the transmit power and the zero/one channel assignment indicator for the  $m^{th}$ user of the  $j^{th}$ cell on the  $n^{th}$ sub-channel. The indicator variable $x_{mj}^{n}$ is one if the  $n^{th}$ sub-channel of the  $j^{th}$ cell is assigned to the  $m^{th}$ user of this cell; otherwise $x_{mj}^{n}=0$. The main objective persuaded in the current network is to allocate equal number of sub-channels to each active link such that the network throughput is maximized, assuming each link is s.t. a peak transmit power constraint. Also, if  $R_{mj}^{n}$ denotes the rate of the $m^{th}$ user in the $j^{th}$ cell on the $n^{th}$ sub-channel, under Gaussian codebook assumption, we have:\\
\begin{eqnarray}\label{sysmod2}
R^{n}_{mj}=\log_2\left(1+\frac{ p^{n}_{mj} h^{nj}_{mj}}{\sigma^2+\sum_{j'\neq j}\sum_{m=1}^{M}x^n_{mj'} p^{n}_{mj'} h^{nj}_{mj'}} \right)
\end{eqnarray}
where $\sigma^2$ is the variance of additive white Gaussian noise (AWGN).

We consider specific quality of service (QoS) requirement for each user called the minimum required data rate which has some applications such that video phone calls and etc. Thus the optimization problem can be cast as the following:
\begin{equation}\label{max_prob.2}
\begin{aligned}
\max_{\textbf{x,p}}\sum_{j=1}^{2}\sum_{m=1}^{M}&\sum_{n=1}^{N}\log_2\left(1+\frac{x^n_{mj}p^{n}_{mj} h^{nj}_{mj}}{\sigma^2+\sum_{j'\neq j}\sum_{m=1}^{M} x^n_{mj'} p^{n}_{mj'} h^{nj}_{mj'}} \right)\\
\text{s.t.}~~ &{{C}_{1}}:~\underset{n=1}{\overset{N}{\mathop \sum }}\,x_{mj}^{n}p_{mj}^{n}\le {{p}_{max}}\\
&{{C}_{2}}:~p_{mj}^{n}\ge 0\\
&{{C}_{3}}:~\underset{n=1}{\overset{N}{\mathop \sum }}\,x_{mj}^{n}R_{mj}^{n}\ge {{R}_{min}}\\
&{{C}_{4}}:~\underset{m=1}{\overset{M}{\mathop \sum }}\,x_{mj}^{n}=1\\
&{{C}_{5}}:~x_{mj}^{n}\in \left\{ 0,1 \right\}\\
\end{aligned}
\end{equation}
where the constraint $C_{1}$ indicates the total transmit power of each user is limited to the $p_{max}$. The constraint $C_{2}$ is guarantying the positivity of allocated power of each user. The constraint $C_{3}$ guarantees the minimum required data rate of each user. The constraint $C_{4}$ ensuring each sub-channel is assigned to only one user, and $C_{5}$ indicates that the sub-channel indices are binary variables. It is noteworthy that since $x^{n}_{mj}$ is a binary variable we can write:
\begin{eqnarray}\label{sysmod3}
x^{n}_{mj}R^{n}_{mj}=\log_2 \left(1+\frac{ x^{n}_{mj}p^{n}_{mj} h^{nj}_{mj}}{\sigma^2+\sum_{j'\neq j}\sum_{m=1}^M x^n_{mj'} p^{n}_{mj'} h^{nj}_{mj'}}  \right)
\end{eqnarray}
It should be noted that when $x^{n}_{mj}=0$, both sides of (3) becomes zero. Similarly, referring to the definition of $R^{n}_{mj}$ in (\ref{sysmod2}), equation (\ref{sysmod3}) yields to (\ref{sysmod2}) for the case of $x^{n}_{mj}=1$.

One can readily verify that the optimization problem in~(\ref{max_prob.2}) involves some continuous variables $p_{mj}^{n}$ as well as integer variables $x_{mj}^{n}$ and so it is not convex in general. Thus, it does not yield to a trivial solution. This motivated us to pursue addressing the aforementioned problem through relying on two approximations which yields to a tight result in high SINR regimes. Then the simplified problem is tackled in two steps:(i) attempting to assign best sub-channel to each link and (ii) finding the best power allocation strategy across sub-channels. 
\section{Sub-channel assignment}
This section tends to determine the sub-channel assignment for each user. It is assumed that all users use all of their available power and operate in high SINR region. So, in this case, the achievable rate
\begin{align}\label{sysmod4}
R^{n}_{mj}=\log_2 \left(1+\frac{ p^{n}_{mj} h^{nj}_{mj}}{\sigma^2+\sum_{j'\neq j}\sum_{m=1}^M x^n_{mj'} p^{n}_{mj'} h^{nj}_{mj'}}  \right)
\end{align}
is much greater than zero. Thus, this term can be approximated as:
\begin{align}\label{sysmod5}
\log_2(1+\text{SINR})&\stackrel{a}{\approx}\log_2(\text{SINR}) \nonumber \\
&\stackrel{b}{\approx}\log_2(\text{SIR})=\log(p^{n}_{mj}h^{nk}_{mj})-\log(p^{n}_{mj'}h^{nj}_{mj'})
\end{align}
where $a$ comes from the fact that $\log_2(1+x)\approx\log_2(x)$ for large values of $x$. Also, $b$ is used because of high SNR assumption and so the noise power is very smaller than interference power. As a result, we can use equation (\ref{sysmod5}) as an approximated achievable rate for $m^{th}$ user in the $j^{th}$ cell on the $n^{th}$ sub-channel. To evaluate this, it is needed to know the best sub-channel in each cell for $m^{th}$ user. Moreover, in the studied network, each user in one of cells has the same sub-channel as a user in another cell. These two users having the same sub-channel named as a user pair. Different user pairs have different sub-channels and their transmitted signals are orthogonal to each other. Therefore, each user pair has no effect on another pairs. Typically, the sum rate for user pair including user-1 and user-2 can be written as:
\begin{align}\label{sysmod6}
R_{11}^{n}+R_{12}^{n}&= \log⁡_2(p_{11}^{n} h_{11}^{n1})-\log_2⁡(p_{12}^{n} h_{12}^{n1})+\log_2⁡(p_{12}^{n} h_{12}^{n2})
\nonumber \\
&~~~~~~- \log⁡_2(p_{11}^{n} h_{11}^{n2})
\nonumber \\
&=\left(\log⁡_2(h_{11}^{n1})+\log_2⁡(h_{12}^{n2})\right)\nonumber \\
&~~~~~~-\left(\log⁡_2(h_{11}^{n2})+\log_2⁡(h_{12}^{n1})\right)
\end{align}
According to (\ref{sysmod6}), the sum rate of each user pair is independent of the transmit power of its users. So, for each user pair, the sum rate can be approximated as a combination of two terms. The first term has a positive impact on the sum rate and related to the direct channel gains. The second term has negative impact and corresponds to the cross channel gains. Therefore, the sum rate of each pair has a benefit and a cost in the throughput of the network.

As a result, the optimization problem can be reformulated as,
\begin{equation}\label{max_prob.3}
\begin{aligned}
&\max_{\textbf{x}}\sum_{j=1}^2\sum_{m=1}^M\sum_{n=1}^N\ x^n_{mj}\log_2 \left(\frac{p^{n}_{mj} h^{nj}_{mj}}{\sum_{j'\neq j}\sum_{m=1}^Mx^n_{mj'} p^{n}_{mj'} h^{nj}_{mj'}} \right)\\
&\text{s.t.}~~ {{C}_{1}}:~~\underset{m=1}{\overset{M}{\mathop \sum }}\,x_{mj}^{n}=1,~~~ n\in\{1,2,...,N\},~ j\in\{1,2\} \\
&~~~~~{{C}_{2}}:~x_{mj}^{n}\in \left\{ 0,1 \right\}
\end{aligned}
\end{equation}
Based on above, the sum rate maximization problem can be tackled through the so called assignment problem. In a typical assignment problem, there are N machines and N jobs, where the cost of doing the $j^{th}$ job on the $m^{th}$ machines is $c_{mj}$. The problem is effectively assign these jobs to available machines such that the total cost is minimized. Let $C^{n}_{mj}$ be the cost of assigning the $n^{th}$  sub-channel to the $m^{th}$ user in the $j^{th}$ cell. We define the $n\times n$ cost matrix C associated with the assignment problem such that the element of the $m^{th}$ row and the $n^{th}$ column is set to
\begin{align}\label{Cost}
    C^{n}_{mj}=log_2(\frac{h_{mj}^{nj} }{ h_{mj}^{nj'}}),~~ m\in\{1,...,M\},n\in\{1,...,N\}, j\in\{1,2\}
\end{align}
The Hungarian method is a classical method to solve this kind of assignment problem. It is a combinatorial optimization algorithm which solves assignment problem in a polynomial time~\cite{b15}. Thus, the Hungarian method can find the best sub-channel assignment which solves the:
\begin{align}\label{max_prob.4_0}
&\min_{\textbf{x}}\sum_{j=1}^2\sum_{m=1}^M\sum_{n=1}^N\ -x^n_{mj}C^{n}_{mj}\nonumber\\
&\text{s.t.}~~ {{C}_{1}}:\underset{m=1}{\overset{M}{\mathop \sum }}\,x_{mj}^{n}=1,~~~ n\in\{1,2,...,N\},~ j\in\{1,2\}\nonumber\\
&~~~~~{{C}_{2}}:~x_{mj}^{n}\in \left\{ 0,1 \right\}
\end{align}
where $x^n_{mj}$  is an integer variable and indicator ensuring each sub-channel is assigned to one user pair based on the constraint $C_{1}$ in (\ref{max_prob.4_0}). It should be noted that the assignment problem can be extended to a more general case of having N sub-channels and K users through without imposing any constraint on the size of sub-channels and users.
The main advantage of this method is having low complexity. 
\section{Power Allocation Strategy }\label{AA}
In the previous section, each sub-channel is assigned to one user. In the studied network, each user in one of cells has the same sub-channel as a user in another cell. These two users having the same sub-channel named as a user pair. Different user pairs have different sub-channels and their transmitted signals are orthogonal to each other. Therefore, each user pair has no effect on another pairs. Thus, the total network throughput maximization problem can be replaced by the throughput maximization in each user pair. Typically, the sum rate maximization problem for user pair including user-1 and user-2 can be written as:
\begin{align}\label{max_prob.3}
\max_{p^{n}_{11},p^{n}_{12}}\log_2(1+&\frac{p_{11}^{n} h_{11}^{n1}}{p_{12}^{n} h_{12}^{n1}+\sigma^2})+\log_2⁡(1+\frac{p_{12}^{n} h_{12}^{n2}}{p_{11}^{n} h_{11}^{n2}+\sigma^2})\nonumber\\
\text{s.t.}~ &{{C}_{1}}:0\le p_{11}^{n}\le {{p}_{max}}\nonumber\\
&{{C}_{2}}:0\le p_{12}^{n}\le {{p}_{max}}\nonumber\\
&{{C}_{3}}:R_{11}^{n}\ge {{R}_{min}}\nonumber\\
&{{C}_{4}}:R_{12}^{n}\ge {{R}_{min}}
\end{align}
For simplicity, some variables are changed using the definitions $a\triangleq h_{11}^{n1}$, $b\triangleq h_{12}^{n1}$, $c\triangleq h_{12}^{n2}$ \text{and} $d\triangleq h_{11}^{n2}$. Then, the above problem is changed as follows:
\begin{align}\label{max_prob.4}
\max_{p^{n}_{11},p^{n}_{12}}\log_2⁡(1+&\frac{p_{11}^{n}a}{p_{12}^{n}b+\sigma^2})+\log_2⁡(1+\frac{p_{12}^{n}c}{p_{11}^{n}d+\sigma^2})\nonumber\\
\text{s.t.}~ &{{C}_{1}}:0\le p_{11}^{n}\le {{p}_{max}}\nonumber\\
&{{C}_{2}}:0\le p_{12}^{n}\le {{p}_{max}}\nonumber\\
&{{C}_{3}}:\log⁡(1+\frac{p_{11}^{n}a}{p_{12}^{n}b+\sigma^2})\ge {{R}_{min}}\nonumber\\
&{{C}_{4}}:\log⁡(1+\frac{p_{12}^{n}c}{p_{11}^{n}d+\sigma^2})\ge {{R}_{min}}
\end{align}
Using some manipulations, the optimization problem (\ref{max_prob.4}) can be replaced by
\begin{equation}\label{max_prob.5}
\begin{aligned}
\max_{p^{n}_{11},p^{n}_{12}}\log_2⁡(1+&\frac{p_{11}^{n}a}{p_{12}^{n}b+\sigma^2})+\log_2⁡(1+\frac{p_{12}^{n}c}{p_{11}^{n}d+\sigma^2})\\
\text{s.t.}~ &{{C}_{1}}:0\le p_{11}^{n}\le {{p}_{max}}\\
&{{C}_{2}}:0\le p_{12}^{n}\le {{p}_{max}}\\
&{{C}_{3}}:p_{11}^{n}\ge \frac{(2^{{R}_{min}}-1)p_{12}^{n}b}{a}+ \frac{(2^{{R}_{min}}-1)\sigma^2}{a} \\
&{{C}_{4}}:p_{12}^{n}\ge \frac{(2^{{R}_{min}}-1)p_{11}^{n}d}{c}+ \frac{(2^{{R}_{min}}-1)\sigma^2}{c} \\
\end{aligned}
\end{equation}
or, equivalently, as,
\begin{equation}\label{max_prob.6}
\begin{aligned}
\max_{p^{n}_{11},p^{n}_{12}}\log_2⁡(1+&\frac{p_{11}^{n}a}{p_{12}^{n}b+\sigma^2})+\log_2⁡(1+\frac{p_{12}^{n}c}{p_{11}^{n}d+\sigma^2})\\
\text{s.t.}~ &{{C}_{1}}:0\le p_{11}^{n}\le {{p}_{max}}\\
&{{C}_{2}}:0\le p_{12}^{n}\le {{p}_{max}}\\
&{{C}_{3}}:p_{11}^{n}\ge \ p_{12}^{n}b{a'}+a'\sigma^2 \\
&{{C}_{4}}:p_{11}^{n}\le \frac{p_{12}^{n}c'}{d}- \frac{\sigma^2}{d} \\
\end{aligned}
\end{equation}
where $a'\triangleq \frac{(2^{Rmin}-1)}{a}$ and $c'\triangleq \frac {c}{(2^{Rmin}-1)}$.

The third and forth constraints denote to two half-spaces. These half-spaces are, respectively, related to lines $l_1: p_{11}^{n} = \ p_{12}^{n}b{a'}+a'\sigma^2$ and $l_2: p_{11}^{n} = \frac{p_{12}^{n}c'}{d}- \frac{\sigma^2}{d}$. Lines $l_1$ and $l_2$ have positive slopes and they cross $p_{11}^{n}\text{-axis}$, respectively, at positive and negative values of $p_{11}^{n}$. Therefore, $C_3$ and $C_4$ can have intersection in the first quadrant of $p_{12}^{n}-p_{11}^{n}$ plane, if and only if the slope of $l_1$ is less than that's of $l_2$, see Fig.1. Also, to satisfy the constraints $C_1$ and $C_2$, it is needed to reside the junction point of $l_1$ and $l_2$ in the first and second constraints. By other words, the optimization problem (\ref{max_prob.6}) is feasible, if the following constraints are satisfied:
\begin{align}\label{feasibility_cond1}
   & ba' < \frac{c'}{d} \\
   0 \leq p_{11,b}^{n} \leq p_{max} &~~~  \text{and}~~~ 0 \leq p_{12,b}^{n} \leq p_{max} \label{feasibility_cond2}
\end{align}
where the point $(p_{11,b}^{n},p_{12,b}^{n})$ refers to the junction point and is obtained as following:
\begin{eqnarray}\label{sysmod10}
p_{12}^{n}b{a'}+a'\sigma^2= \frac{p_{12}^{n}c'}{d}-\frac{\sigma^2}{d}\Rightarrow p_{12,b}^{n}=\frac{\frac{\sigma^2}{d}+a'\sigma^2}{\frac{c'}{d}-ba'}\\
p_{11,b}^{n}= \frac{p_{12,b}^{n}c'}{d}-\frac{\sigma^2}{d}\Rightarrow p_{11,b}^{n}=\frac{\frac{\sigma^2}{d}+a'\sigma^2}{\frac{c'}{d}-ba'}{\frac{c'}{d}}-\frac{\sigma^2}{d}
\end{eqnarray}
\begin{figure} \label{fig:fig_Feasibility}
  \centering
  \includegraphics[width=8.00cm]{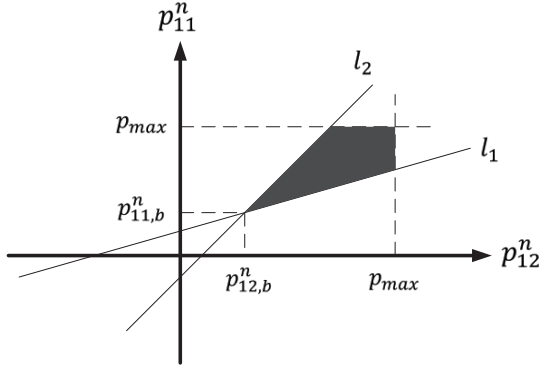}
  \caption{Feasibile set of power allocation problem}
\end{figure}
The above explanations about the feasibility condition of (\ref{max_prob.6}) are illustrated in Fig.1 and the feasible set is shown by black region. As seen, if the slope of $l_1$ is greater than that's of $l_2$, the intersection of mentioned half-spaces don't reside in the first quadrant of plane. Also, if the constraint (\ref{feasibility_cond2}) isn't held, this intersection region cannot satisfy the power constraints.

When the constraints (\ref{feasibility_cond1}) and (\ref{feasibility_cond2}) are satisfied, the optimization problem (\ref{max_prob.6}) is feasible. Otherwise, this problem cannot be solved and so all users are set in "off" state. In the feasible case, to solve (\ref{max_prob.6}), the sign of the derivative of the following function is explored.
\begin{eqnarray}\label{sysmod11}
f(p^{n}_{11},p^{n}_{12})=\log⁡(1+\frac{p_{11}^{n}a}{p_{12}^{n}b+\sigma^2})+\log⁡(1+\frac{p_{12}^{n}c}{p_{11}^{n}d+\sigma^2})
\end{eqnarray}
At first, the variable $p_{12}^{n}$ is assumed as a constant value. The derivative of $f(p_{11}^{n},p_{12}^{n} )$ can be written as,
\begin{eqnarray}\label{sysmod12}
\frac{\partial f(p_{11}^{n},p_{12}^{n})}{\partial p_{11}^{n}}=\frac{\frac{a}{p_{12}^{n}b+\sigma^2}}{1+\frac{{p_{11}^{n}}a}{{p_{12}^{n}}b+\sigma^2}}-\frac{\frac{p_{12}^{n}cd}{(p_{11}^{n}d+\sigma^2)^2}}{1+\frac{p_{12}^{n}c}{p_{11}^{n}d+\sigma^2}}=0
\end{eqnarray}
or, equivalently, as
\begin{eqnarray}\label{sysmod13}
 \frac{a}{p_{12}^{n}b+{p_{11}^{n}}a+\sigma^2}- \frac{p_{12}^{n}cd}{({p_{11}^{n}d+}p_{12}^{n}c+\sigma^2)(p_{11}^{n}d+\sigma^2)}=0
\end{eqnarray}
or,
\begin{eqnarray}\label{sysmod14}
ad^{2}(p_{11}^{n})^2+2ad\sigma^2 p_{11}^{n}+a\sigma^4+p_{12}^{n}ac\sigma^2\nonumber\\
-bcd(p_{12}^{n})^2-cd\sigma^2p_{12}^{n}=0
\end{eqnarray}
The problem (\ref{sysmod14}) is a quadratic problem and its roots are given by,
\[ p_{11}^{n}=
  \begin{cases}
  -\sigma^2/d\pm\frac{\sqrt{p_{12}^{n}acd^2(d\sigma^2-a\sigma^2+bdp_{12}^{n})}}{ad^2}      & \quad \text{if } \triangle > 0 \text{}\\
   \text{No real roots}  & \quad \text{if } \triangle < 0\text{}\\
  \frac{-\sigma^2}{d}           & \quad \text{if } \triangle = 0\text{}\\
  \end{cases}
\]
where $\triangle=\frac{\sqrt{p_{12}^{n}acd^2(d\sigma^2-a\sigma^2+p_{12}^{n}acd^2)}}{ad^2}$.

In proceeding, to explore the sign of $\frac{\partial R}{\partial p_{11}^{n} }$, each cases of the above equation should be studied.
\begin{itemize}
\item Case~1: If $\triangle>0$, $\frac{\partial R}{\partial p_{11}^{n}} $ has two roots. The lowest and greatest roots are, respectively, denoted by $p_{{11}_{1}}^{n}$ and $p_{{11}_{2}}^{n}$. $\frac{\partial R}{\partial p_{11}^{n}} $ is negative between two roots and is positive for otherwise case, and so the function $f(p_{11}^{n},p_{12}^{n} )$ has decreasing or increasing characteristics, respectively, in these cases. Therefore, we can arrive at followings for studied region:
    \begin{align}\label{fff}
    &\text{If}~~~ p_{{11}_{2}}^{n}\leq p_{{11,b}}^{n}\nonumber\\
    & \hspace{5mm}   p_{11}^{n^{*}}(p_{12}^{n})=p_{1_{max}}\nonumber\\
    &\text{Else if}~~~    p_{11,b}^{n}\leq p_{{11}_{2}}^{n}\leq p_{1_{max}} \nonumber\\
    & \hspace{5mm}   p_{11}^{n^{*}}(p_{12}^{n})=\text{arg~max}\Big( f(p_{11,b}^{n},p_{12}^{n}),f(p_{1_{max}}^{n},p_{12}^{n})\Big)\nonumber\\
    &\text{otherwise}~~~  p_{11}^{n^{*}}(p_{12}^{n})=p_{11,b}^{n}
    %&p_{11}^{n^{*}}(p_{12}^{n})=
%        \begin{cases}
%        p_{1_{max}} ~~~~ \text{if}~~ p_{{11}_{2}}^{n}\leq p_{{11}_{bott}}^{n}\text{}\\
%        \text{arg~max}(f(p_{11,b}^{n},p_{12}^{n}),f(p_{1_{max}}^{n},p_{12}^{n})) \\
%        ~~~~~~~ \text{if }~~p_{11,b}^{n}\leq p_{{11}_{2}}^{n}\leq p_{1_{max}}\text{}\\
%        p_{11,b}^{n} ~~~~ \text{otherwise }
%        \end{cases}
    \end{align}
where $p_{11}^{n^*}(p_{12}^{n})$ is the optimal value of $(p_{11}^{n})$ for each value of $p_{12}^{n}$.
\item Case~2: If $\triangle <0 ~\text{or} ~\triangle=0$.
In these two cases, the function $\frac{\partial R}{\partial p_{11}^{n}}$ is always positive and so $p_{11}^{n^*}(p_{12}^n )=p_{1_{max}}$.
\end{itemize}
\begin{remark}\label{remark1}
    According to the results of case {1} and case {2} for each value of $p_{12}^{n}$, the optimal value of $p_{11}^{n}$ is either $p_{11,b}^{n}$ or $p_{1_{max}}$.
\end{remark}
To find the optimal value of $p_{12}^{n}$, the same procedure, as $p_{11}^{n}$, is done. To do this, the variable $p_{11}^{n}$ is assumed as a constant value. The derivative of $f(p_{11}^{n},p_{12}^{n})$ with respect to $p_{12}^{n}$ can be written as:
\begin{eqnarray}\label{sysmod16}
    \frac{\partial f(p_{11}^{n},p_{12}^{n})}{\partial p_{12}^{n}}=\frac{\frac{c}{p_{11}^{n}d+\sigma^2}}{1+\frac{{p_{12}^{n}}c}{{p_{11}^{n}}d+\sigma^2}}-\frac{\frac{p_{11}^{n}ab}{(p_{12}^{n}b+\sigma^2)^2}}{1+\frac{p_{11}^{n}a}{p_{12}^{n}d+\sigma^2}}=0
\end{eqnarray}
\begin{remark}\label{remark2}
  It is seen that the equations (\ref{sysmod12}) and (\ref{sysmod16}) have the same mathematical form and so the optimal value of $p_{12}^{n}$ as a function of $p_{11}^{n}$, i.e., $p_{12}^{n^*}(p_{11}^{n} )$ is taken at $p_{12,b}^{n}$ or $p_{2_{max}}$.
\end{remark}
\begin{proposition}
  Based on remark~\ref{remark1} and remark~\ref{remark2}, the optimum value of the pair $(p_{11}^{n},p_{12}^{n})$, i.e.,$(p_{11}^{n^*},p_{12}^{n^*})$, is occurred at one of the following points:
  \begin{eqnarray}
  (p_{11}^{n^*},p_{12}^{n^*})=
  \begin{cases}
    (p_{11,b}^{n},p_{12,b}^{n})\\
    (p_{11,b}^{n},p_{2_{max}})\\
    (p_{1_{max}},p_{12,b}^{n})\\
    (p_{1_{max}},p_{2_{max}})\\
  \end{cases}
  \end{eqnarray}
\end{proposition}
Therefore, to find the optimal powers, for each user pair, a 4-point search is done and each point which yields to a highest sum rate is selected. Thus, it is seen that the optimal power allocation is provided as an analytical solution and so this method is an efficient method from complexity viewpoint. 
\section{Simulation Result}
In this section, we investigate the performance of proposed algorithm used for resource allocation and power control in the uplink direction of a two-cell network. It is assumed that there are two cells and each cell has three users. Six users are placed uniformly in the coverage area of network and $N=3$ orthogonal sub-channels are shared in each cell. The wireless channel model is Rayleigh flat fading and it is modeled as $\phi^n d_{i(j),j'}^{-\alpha}$ dB where the distance d is in units of meter. In this model, $\phi^n$ is a random variable generated according to the Rayleigh distribution with unit variance. Moreover, the path-loss effect is modeled by $d_{i(j),j'}^{-\alpha}$ where $\alpha=3$ is the path loss exponent. The distance between each user and corresponding base station is assumed as $100m$ and also the distance between it and another base station is assumed as $500m$. To illustrate the performance of proposed algorithm and compare it with existing work, many channel realizations are generated and the optimization problem is solved for each realization. Then, the average of network throughput is plotted versus the signal-to-noise ratio.
\begin{figure} \label{fig:fig1}
  \centering
  \includegraphics[width=8.00cm,height=6.00cm]{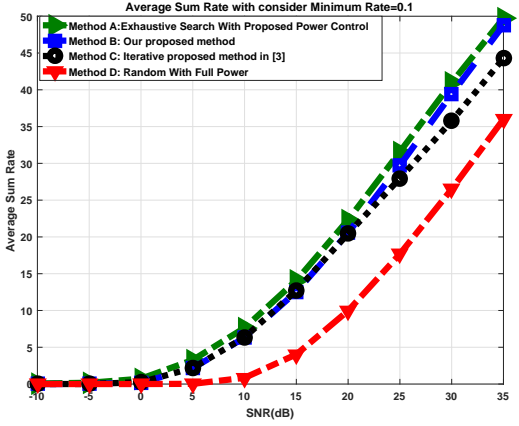}
  \caption{Sum rate when $R_{min}=0.1$ and variance noise=-110db for different values of $p_{max}$}
\end{figure}

Fig.2 illustrates the average sum rate versus the SNR for different methods.
 Method A represents  the optimal solution for which channel assignment is performed using exhaustive search and the optimal power control is done according to our proposed optimal power allocation policy.For this curve ,all of possible sub-channel assignment are considered and the optimum power allocation, given by this paper.
 Method B depicts the proposed method for sub-channel assignment based on the Hungarian method and the proposed optimal power control ,is computed for each of them.
 Method C, the channel assignment and power allocation are obtained using the proposed method in \cite{b3}. Finally,Method D, indicates the case in which channel assignment is done randomly and all users transmit their signals with full-power. In each scenario, if the rate of each user dose not satisfy the minimum required rate, the optimization problem is not feasible and so all users are set in "off" state.
As observed in Fig.2, in high SNR region, the proposed method is near to optimal method and so it can be used as a close to optimal solution in this SNR region. Moreover, it is seen that the our method has not significant degradation in low SNRs from optimal solution. According to the fact that the proposed method is more simpler than the exhaustive search and has less complexity, this method is provided as a good alternative in all range of SNRs. Moreover, it is shown that in the current work gives higher throughput than \cite{b3} in the wide range of SNRs.

\begin{figure} \label{fig:fig2}
  \centering
  \includegraphics[width=8.00cm]{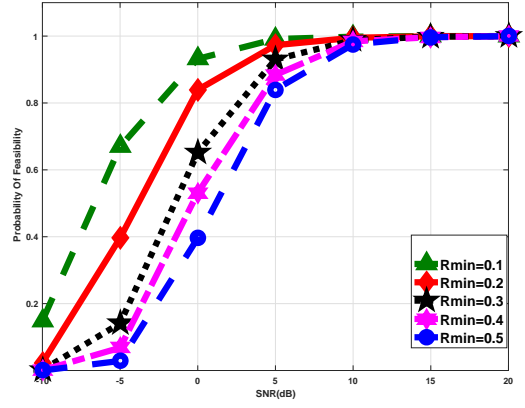}
  \caption{Probability of feasibility for different values of $R_{min}$ when $p_{max}=30db$ and $\sigma^{2}=-110db$ }
\end{figure}

\begin{figure}\label{fig:fig3}
  \centering
  \includegraphics[width=8.00cm]{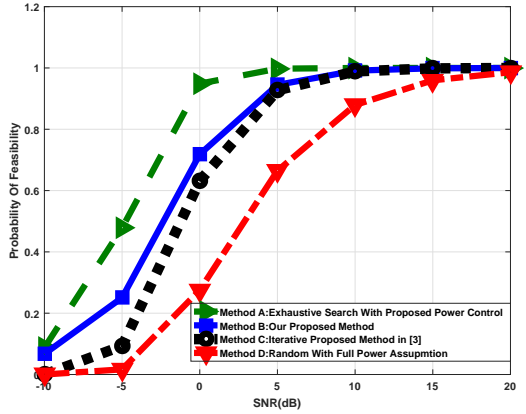}
  \caption{Probability of feasibility for different methods when $R_{min=0.1}$, $p_{max}=30db$ and $\sigma^{2}=-110db$ }
\end{figure}

Fig.3 shows the probability of feasibility for different values of $R_{min}$ when the proposed method is used. This figure represents that the probability of feasibility is increased by increasing SNR. Also, it is seen that decreasing $R_{min}$ yields the feasible set of optimization problem becomes larger and so the feasibility probability is increased.

Fig.4 illustrate the probability of feasibility for different methods when $R_{min}$ is set to $0.1$. As previous plot, increasing SNRs yields increasing the probability of feasibility in all methods. Moreover, in low SNRs, the feasibility probability of exhaustive search is greater than the proposed method. But, in high SNR region, the feasibility probability of proposed method ia approximately coincides the optimal method.

\section{Conclusion}\label{sec:conc}
This paper concerns maximizing sum rate for uplink direction the two-cell network. The analytical optimal solution of this non-convex mix-integer nonlinear problem is found to be computationally impossible. So it is proposed to reformulate the optimization problem to a mathematically tractable form. By rewrite the objective function for high SNR region, a Hungarian method is used to find the best sub-channel and then the transmit power of users is determined in analytical form. Simulation results show that our proposed method outperforms another solutions and tends to optimal strategies for high SNRs. 

%\begin{table}[htbp]
%\caption{Table Type Styles}
%\begin{center}
%\begin{tabular}{|c|c|c|c|}
%\hline
%\textbf{Table}&\multicolumn{3}{|c|}{\textbf{Table Column Head}} \\
%\cline{2-4}
%\textbf{Head} & \textbf{\textit{Table column subhead}}& \textbf{\textit{Subhead}}& \textbf{\textit{Subhead}} \\
%\hline
%copy& More table copy$^{\mathrm{a}}$& &  \\
%\hline
%\multicolumn{4}{l}{$^{\mathrm{a}}$Sample of a Table footnote.}
%\end{tabular}
%\label{tab1}
%\end{center}
%\end{table}
%
%\begin{figure}[htbp]
%\centerline{\includegraphics{fig1.png}}
%\caption{Example of a figure caption.}
%\label{fig}
%\end{figure}

\end{document}